# ASTRONOMICAL ORIENTATIONS IN SANCTUARIES OF DAUNIA[1]

E. ANTONELLO[1], V.F. POLCARO[2], A.M. TUNZI SISTO[3], M. LO ZUPONE[3]
[1]INAF-Osservatorio Astronomico di Brera;  e-mail: elio.antonello@brera.inaf.it
[2]INAF-IASF, Rome
[3]Soprintendenza per i Beni Archeologici della Puglia

**Abstract:** Prehistoric sanctuaries of Daunia date back several thousand years. During the Neolithic and Bronze Age the farmers in that region dug hypogea and holes whose characteristics suggest a ritual use. In the present note we summarize the results of the astronomical analysis of the orientation of the row holes in three different sites, and we point out the possible use of the setting of the stars of Centaurus. An interesting archaeological confirmation of an archaeoastronomical prediction is also reported.

Daunia is the ancient name of a region in the north of Apulia. This region is very rich of archaeological sites, dating from Neolithic to roman and medieval epoch. Beginning from the 5th millennium BCE the farmers living in the plain of Daunia dug hypogea for ritual purposes, and they were utilized later as burial-grounds. Such sites could be called sanctuaries; some of them are characterized also by long rows of holes.

### First stage: 2007, Madonna di Loreto, Trinitapoli

In the Bronze Age sanctuary of Madonna di Loreto in Trinitapoli there are some interesting hypogea that were dug in the thick stratum of calcareous rock (called *crusta*) located below the soil, but the most impressive phenomenon is the rows with hundred holes. They were dug from 18th to 9th century BCE, that is during about one thousand years. The prehistoric holes have roughly a circular shape, and can be easily discriminated from the rectangular shallow groves that were excavated in past centuries for farming (such as artichoke cultivation). The characteristics of the holes and the remains found inside indicate a ritual use, while other utilizations should be excluded. They are not post holes and cannot be justified with cultivation (e.g. for trees and plants).

The discovery of the holes was not recent, but the archaeologists had not yet considered the regularity of the alignments of the hole rows. In 2007 they suspected a possible astronomical orientation of the rows, and therefore they contacted the astronomers for an analysis. The results of the study were presented at the SEAC 2008 meeting (Tunzi et al., 2009). We assumed that one or few holes per year were dug. It seems that the ancestors adopted three main directions based on the Sun and Moon. The most frequent azimuth coincides within the statistical error with the meridian. The second azimuth, 61.5 degrees, corresponds to the sunrise azimuth for a date that is not far from the summer solstice sunrise. The third azimuth, 129.6 degrees, is that of the Major Lunar Standstill, and the smaller number of the holes in such rows would be compatible with an event that occurred more rarely.

Trinitapoli could be considered the first stage of a journey that is still going on, and is becoming more and more exciting.

### Second stage: 2009, Ponterotto, Ordona

The second stage of the journey occurred in 2009. A new sanctuary with Neolithic hypogea was discovered in Ordona, about thirty kilometres west of Trinitapoli. The occasion was the installation of a cable connecting the wind turbines of a wind farm. Only two areas were excavated, located rather far apart (about 900 meters). Straight rows of holes dug rather carefully in the calcareous stratum were found in both areas. The separation between the holes is generally less than one meter, and that between the rows more than three meters. The archaeologists suspect that the area with the holes has a very large extent. The holes contained just soil, and for the present there is no reliable estimate of their age. We gave a preliminary interpretation based on the setting of the stars of Centaurus-Crux group, and we discussed qualitatively this case at the SEAC 2010 meeting in Gilching, pointing out that an archaeological dating was needed before drawing any reliable conclusion (Antonello et al., 2010).

---

[1] Paper presented at the SEAC 2011 meeting in Evora (Portugal); to appear in the Proceedings, *Stars and stones*, F. Pimenta et al. (eds.)

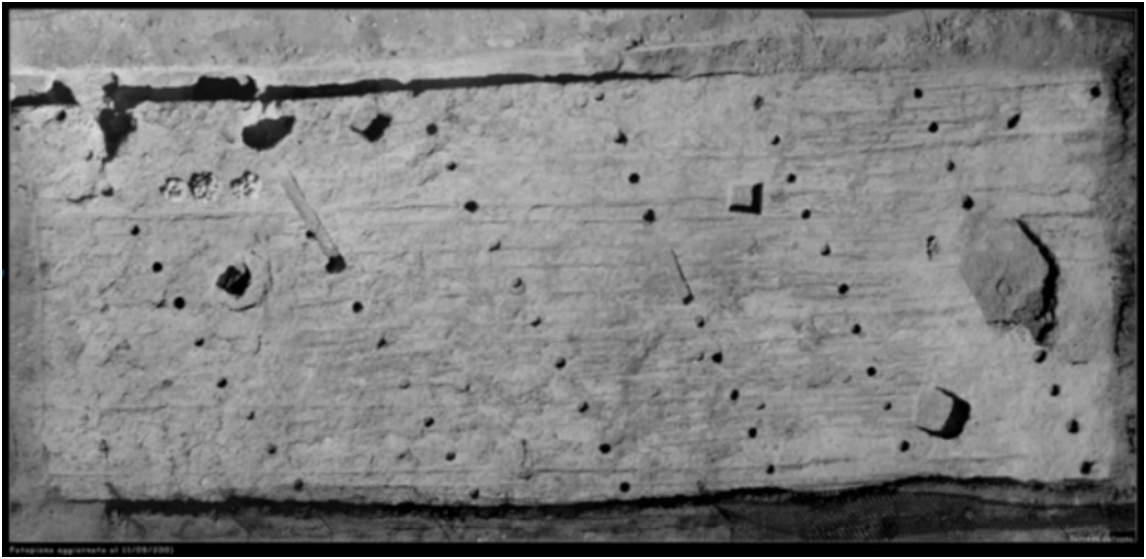

**Fig. 1.** An ortophoto of the site of Ordona – Ponterotto 1. The distance between the holes is less than one meter, and the separation between the rows is about three meters.

One should note a sort of a fan shape of the rows in Figure 1 due to a slightly progressive difference of their orientation. The separation between the rows increases in the upper (roughly north-east; see Figure 2) direction. According to the interpretation based on the setting of Centaurus-Crux, their epoch would be roughly from the mid of second millennium BCE to the end of that millennium, that is from about fifteen hundred years BCE to one thousand years BCE. We assume that the ancestors operated in the following way. They fixed a south-west direction by looking at the setting of a star, for example α Centauri, using a reference, may be a pole. Then they started to dig the holes in the opposite (north-east) direction, just one or few holes per year, and continued for many years as far as the space was available. Then a new row was dug, but with a slightly different direction, since in the meantime the star setting position was changed due to the precession.

The results of an accurate least squares analysis are shown in Figure 2, along with the estimated epochs of the rows, assuming their direction was indicating the setting of α Centauri. The azimuth range is from 194.5 to 203.6 degrees, with a statistical error of about 0.1 – 0.3 degrees. Note that the value of the row on the right side in Figure 2 does not follow the progression, since one would expect a value around 204 degrees. We will come back later on this point.

**Third stage: 2010, Mandriglia, Trinitapoli**

More recently, archaeologists found other holes in several construction sites in Trinitapoli. In particular at Mandriglia, which is located at about one kilometre and half from Madonna di Loreto. The characteristics of the holes, according to the archaeologists, were similar to those of Ordona.

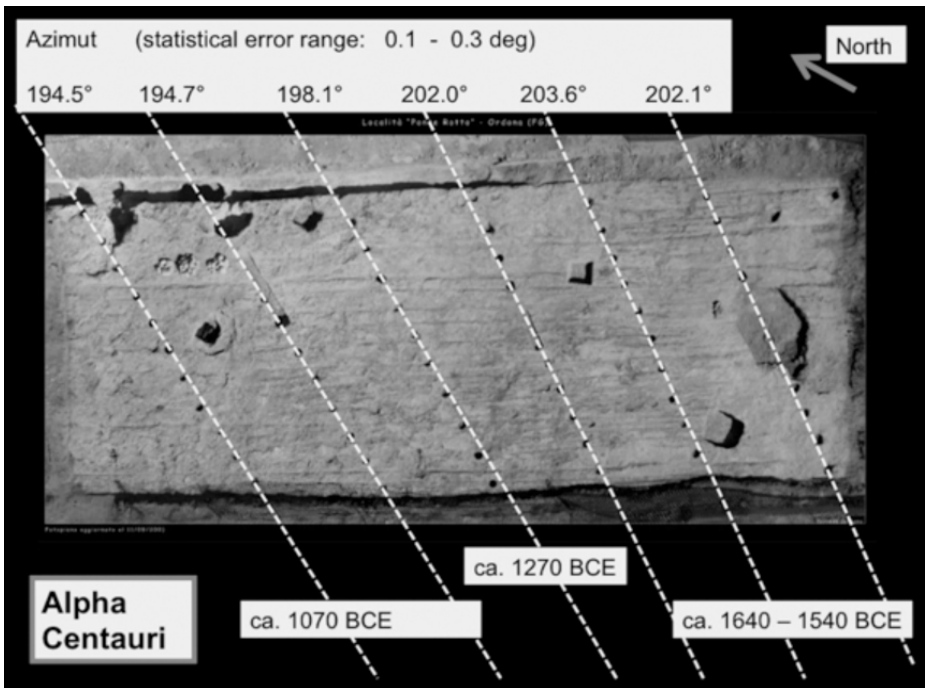

**Fig. 2.** The azimuths of the hole rows of Ordona-Ponterotto 1 and the corresponding approximated epochs of the setting of α Centauri.

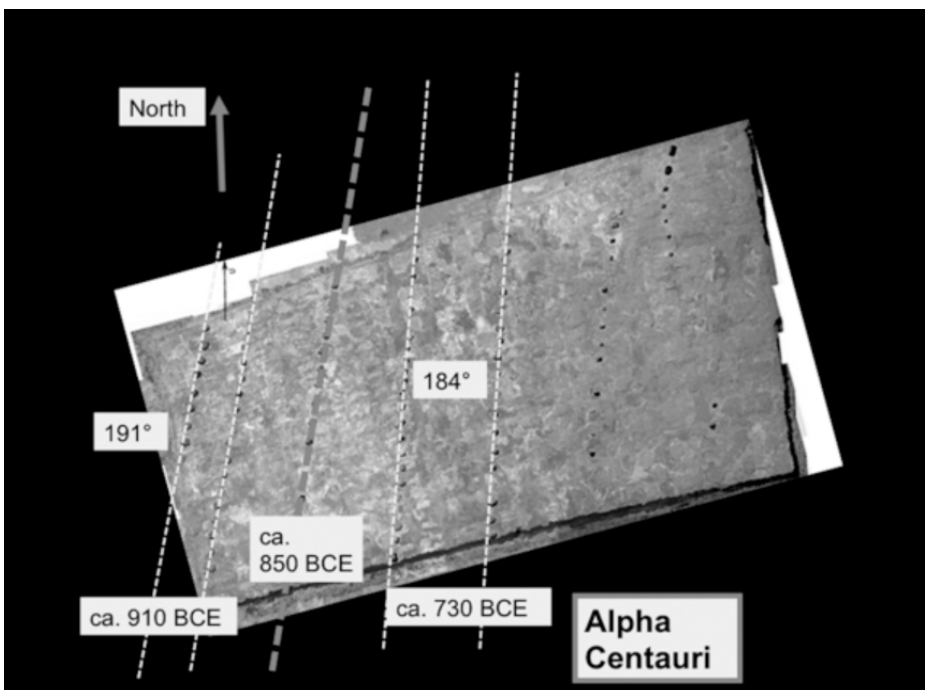

**Fig. 3**. The azimuths of the rows of Mandriglia-Trinitapoli, and the corresponding approximated epochs of the setting of α Centauri. The astronomical epoch of the row marked in grey (ca. 850 BCE) has been confirmed by an archaeological find (9th – 8th century BCE).

There is a more or less progressive change of the orientation, even closer to the southern direction than in the case of Ordona. However, the fan shape is different, since the separation between the rows increases in the southern direction. At the beginning we wondered whether we had been wrong when considering an astronomical orientation based on the stars. The possible explanation of the difference, however, turned out to be very simple in the context of the star setting interpretation.

In the case of Ordona the oldest rows should be those on the right side (east), while in the case of Mandriglia should be those on the left side (west).

Hence, assuming the setting of α Centauri, the epochs for Mandriglia should be those shown in Figure 3. But what about the other holes on the right side? May be the ancestors had increasing difficulties to observe the star since it was very low; we will discuss this point in the next section. Note that the astronomical analysis was performed with no previous knowledge of a possible archaeological dating of the holes, since it was presumed that, as in the case of Ordona, they contained just soil and no archaeological finds. The results of the analysis were presented at the end of October 2010 at a meeting of the Italian Society for Archaeoastronomy (SIA). Also the archaeologists were attending at the meeting and at the end of the presentation, with much surprise, they said they actually found a piece of pottery in one of the holes. The archaeological dating is 9th – 8th century BCE, which corresponds nicely to the astronomical prediction, 850 BCE.

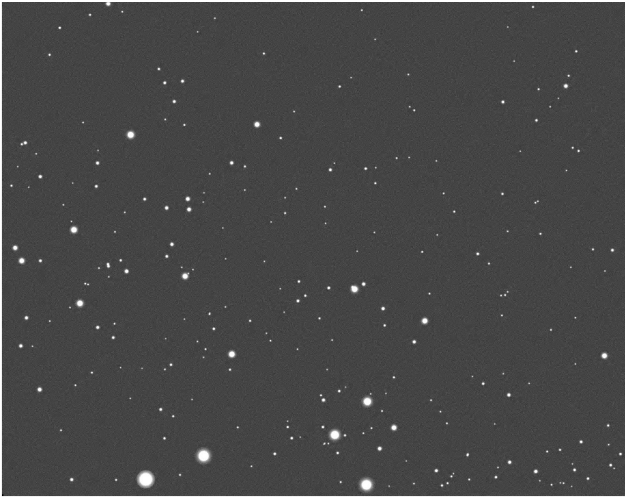

**Fig. 4**. A simulation of the sky in the region of Lupus, Centaurus, Crux, Vela. The lower border represents the horizon. One can see the Southern Cross and the two bright stars, α Centauri and β Centauri, close to the horizon.

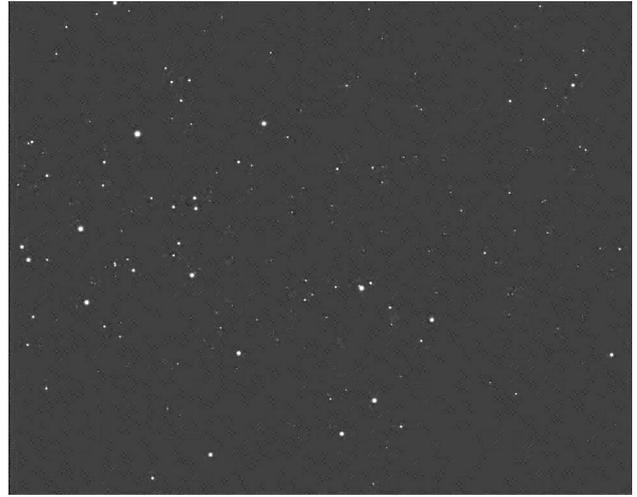

**Fig. 5.** The same as Figure 4 but with the extinction included. α Centauri can be barely detected (it is about two degrees above the horizon).

**Astronomical analysis**

Given the possible importance of this result, let us discuss some astronomical details. We have estimated the air mass according to Young (1994), and the adopted atmospheric extinction coefficient was for a low elevation and for winter time (Schaefer, 1993). We took into account the refraction (Saemundsson, 1986), and also the increased sky background above the horizon (the sunlight is scattered by the atmospheric dust; Garstang, 1989; Leinert et al., 1998). Just to give a feeling to the reader, a simulation of the sky is shown in Figure 4 (with no extinction) and Figure 5 (with extinction included). One can recognize in Figure 4 the Southern Cross, and α Centauri and β Centauri near the horizon. Finally, we assumed the usual limiting value of 6 mag for the visibility of a star. Note that α Centauri is a double star, and moreover its magnitude changes since it is approaching us; however such an effect is small.

The plot in Fig. 6 shows the azimuth of the setting of α Centauri, as seen from Daunia region, against the epoch, along with the orientation of the rows in the archaeological sites.

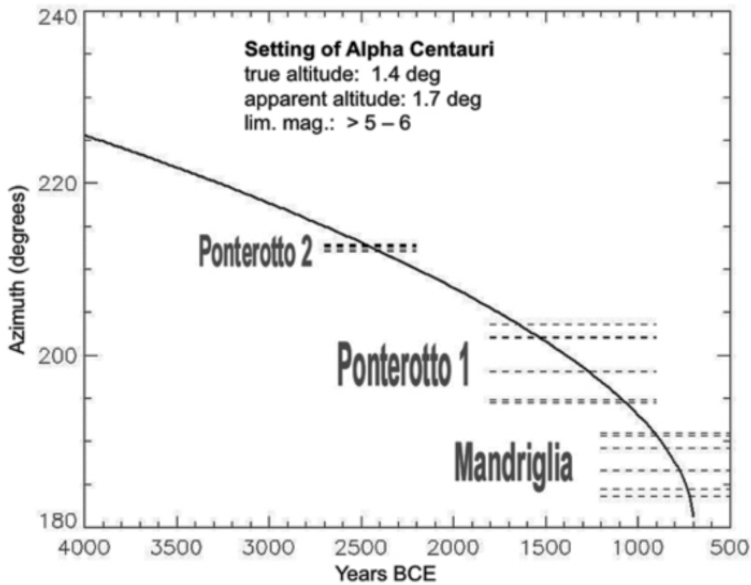

**Fig. 6.** The setting of α Centauri, as seen from Daunia region, plotted against the epoch, and orientation of the rows in the archaeological sites.

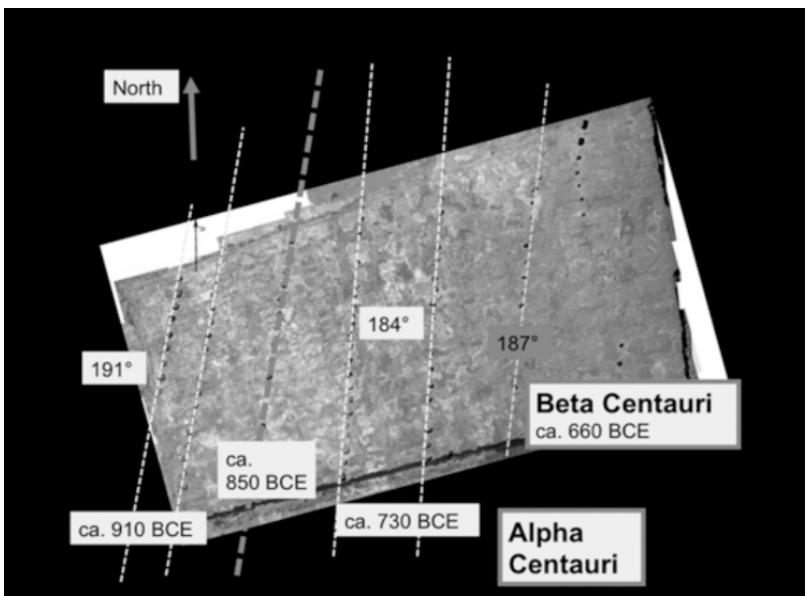

**Fig. 7.** The azimuths of the rows of Mandriglia-Trinitapoli, and the corresponding approximated epochs of the setting of α Centauri and β Centauri.

The star was no more visible around 700 BCE, but β Centauri could be observed, while at a later time, about the 6th century BCE, also β Centauri was no more visible. We think that the ancestors used β Centauri instead of α Centauri during about one century, as long as it was visible. In Figure 7 we show the final results for Mandriglia concerning both α and β Centauri. The result is a bit impressive, because the observed azimuth of the rows put a strong constraint on the possible stars. It is not easy to find two bright stars at the right place in the right time other than α and β Centauri. The last few holes on the right side in Figure 7 could be explained by the fact that, by the 6th century BCE, the ancestors were no more able to observe the traditional star and probably they did not know what to do.

Let us come back to the Ordona case. We should try to justify the small azimuth of the row on the right side of Figure 2. One would expect a value around 204 degrees; that is, there is a difference of about 2 degrees. We could suspect that there were different observing conditions that lasted several years. The extinction was probably larger than in the previous epochs by about half a magnitude, and the limiting magnitude could be observed only at a larger altitude by about half a degree. This could explain reasonably the observed smaller azimuth by about two degrees. A larger extinction can be caused for example by the increased dust in the stratosphere, and the dust could be produced by a very strong volcanic eruption. In the past two centuries this phenomenon has been observed during the strongest eruptions, with the dust remaining in the stratosphere for years, such as in the case of the eruption of Mount Tambora (Indonesia) in 1815. A very strong eruption during the Bronze Age was that of Thera (Santorini, Greece). According to radiocarbon studies, its date would be 1660-1613 BCE (Manning et al., 2006) or 1627-1600 BCE (Friedrich et al., 2006), and one can see that our estimate is compatible with such dates. In the first case the analysis was performed of groups of seeds found in prehistoric storage containers. In the second case, it was based on the rings of an olive branch buried alive in tephra of Santorini. The eruption spread a huge fan of volcanic ash deposits over the Eastern Mediterranean region, and worldwide effects have been ascribed to the eruption: sulphuric acid and fine ash particles in the Greenland Ice Sheet, climatic disturbances in China, and frost damage to trees in Ireland and California (Friedrich et al., 2006; see however Manning et al., 2006).

Among the southern brightest stars, apart from those of Centaurus-Crux, there could be α Piscis Austrinus (Fomalhaut) as a possible target, while there are no northern bright star risings with azimuth around 10 - 20 degrees (i.e. in the opposite direction of the rows). However, in the past seven millennia the minimum azimuth of Fomalhaut was about 197 degrees; therefore this star cannot explain all the observed values. Our present feeling is that, if the setting of bright stars had been used as target, then the stars of Centaurus would be the only solution.

**Conclusion**

The setting of the stars of Centaurus allows a plausible interpretation of the orientations of the rows of Ordona and Mandriglia. The possible significance of such stars for the prehistoric populations of the Mediterranean basin was already pointed out in the past. Hoskin (2001) discussed the prehistoric sanctuaries of Malta (Ggantija temples, about 3500 BCE), Menorca (taula sanctuaries, about 1000 BCE) and the impressive case of the Son Mas sanctuary (about 2000 BCE) of Mallorca. Moreover, during 19th century, astronomers and scholars remarked the spectacular sight of this region of the southern sky, and the unusual twilight effect produced by the brightness of its stars. Such remarkable sight and effect were used by Schiaparelli (1903) to give a reasonable interpretation of a verse in the Bible (Job 9, 9).

Our result is potentially very important, and therefore many archaeological confirmations are required: one archaeological confirmation is not enough. So, we think it is better for the present to consider our interpretation just a fantasy.

Why could it be important? As far as we know, this would be one of the very few places in the world where the precession effect was, so to speak, printed on the ground three thousand years ago. That time, may be a smart person could have realized that something was wrong with the sky; but how to convince others about this idea? It would have been considered a foolish thought of a smart but weird person. We could recall the book of Santillana and von Dekend (1969), where the authors took into account the myths and folk tales of several populations, and tried to prove that the precession effect was already foreshadowed in the myth long time ago. In these myths, Hamlet (Amlodhi) and other similar smart princes feigned dullness (to conceal their intelligence to ensure their safety), while their fabled mill, intended as the rotating sky, had to be wrecked or unhinged. May be the authors of *Hamlet's Mill* had good intuition. They wrote that even if Amlodhi's Quern and the other myths cannot be traced back beyond the Middle Ages, they are derived in different ways from the great and durable patrimony of astronomical tradition, the Middle East (i.e. the Babylonian astronomy, one thousand years before Hipparcus).

In any case, we think that the present work is a good example of the collaboration between astronomers and archaeologists that is going on Italy, and we hope to present at the next SEAC meeting another step of this exciting journey.